\documentclass[pdflatex,sn-nature,twocolumn]{sn-jnl}% Style for submissions to Nature Portfolio journals

\usepackage{graphicx}%
\usepackage{multirow}%
\usepackage{amsmath,amssymb,amsfonts}%
\usepackage{amsthm}%
\usepackage{mathrsfs}%
\usepackage[title]{appendix}%
\usepackage{xcolor}%
\usepackage{textcomp}%
\usepackage{manyfoot}%
\usepackage{booktabs}%
\usepackage{algorithm}%
\usepackage{algorithmicx}%
\usepackage{algpseudocode}%
\usepackage{listings}%
\begin{document}
% comment to exclude from word count
%TC:ignore
\title[Article Title]{A scalable photonic quantum interconnect platform}

\author*[1]{\fnm{Daniel} \sur{Riedel}}\email{daniel.riedel@ionq.co}
\author[2]{\fnm{Teodoro} \sur{Graziosi}}
\author[3]{\fnm{Zhuoxian} \sur{Wang}}
\author[1]{\fnm{Chawina} \sur{De-Eknamkul}}
\author[1]{\fnm{Alex} \sur{Abulnaga}}
\author[1]{\fnm{Jonathan} \sur{Dietz}}
\author[1]{\fnm{Andrea} \sur{Mucchietto}}
\author[1]{\fnm{Michael} \sur{Haas}}
\author[1]{\fnm{Madison} \sur{Sutula}}
\author[1]{\fnm{Pierre} \sur{Barral}}
\author[1]{\fnm{Matteo} \sur{Pompili}}
\author[3]{\fnm{Mouktik} \sur{Raha}}
\author[1]{\fnm{Carsten} \sur{Robens}}
\author[1]{\fnm{Jeonghoon} \sur{Ha}}
\author[1]{\fnm{Denis} \sur{Sukachev}}
\author[1]{\fnm{David} \sur{Levonian}}
\author[1]{\fnm{Mihir} \sur{Bhaskar}}
\author[2]{\fnm{Matthew} \sur{Markham}}
\author*[1]{\fnm{Bartholomeus} \sur{Machielse}}\email{bart.machielse@ionq.co}

\affil[1]{\orgname{IonQ Inc.}, \orgaddress{\street{4505 Campus Dr.}, \city{College Park}, \postcode{20740}, \state{MD}, \country{USA}}}

% \affil[1]{\orgname{IonQ Inc.}, \orgaddress{\street{4505 Campus Dr}, \city{College Park}, \postcode{20740}, \state{MD}, \country{USA}}}

\affil[2]{\orgname{Element Six (UK) Limited, Global Innovation Centre}, \orgaddress{\street{Fermi Avenue, Harwell}, \city{Didcot}, \postcode{OX11 0QR}, \country{UK}}}

\affil[3]{\orgname{AWS Center for Quantum Computing}, \orgaddress{ \city{Pasadena}, \postcode{91006}, \state{CA}, \country{USA}}}

\abstract{Many quantum networking applications require efficient photonic interfaces to quantum memories which can be produced at scale and with high yield. Synthetic diamond offers unique potential for the implementation of this technology as it hosts color centers which retain coherent optical interfaces and long spin coherence times in nanophotonic structures. Here, we report a technique enabling wafer-scale processing of thin-film diamond that combines ion implantation and membrane liftoff, high-quality overgrowth, targeted color center implantation, and serial, high-throughput thermocompression bonding with yields approaching unity. The deterministic deposition of thin diamond membranes onto semiconductor substrates facilitates consistent integration of photonic crystal cavities with silicon-vacancy (SiV) quantum memories. We demonstrate reliable, strong coupling of SiVs to photons with cooperativities approaching 100. Furthermore, we show that photonic crystal cavities can be reliably fabricated across several membranes bonded to the same handling chip. Our platform enables modular fabrication where the photonic layer can be integrated with functionalized substrates featuring electronic control lines such as coplanar waveguides for microwave delivery.
Finally, we implement passive optical packaging with sub-decibel insertion loss. Together, these advances pave the way to the scalable assembly of optically addressable quantum memory arrays which are a key building block for modular photonic quantum interconnects.}
%TC:endignore

\maketitle
\begin{figure*}[h]
\centering
\includegraphics[width=\textwidth]{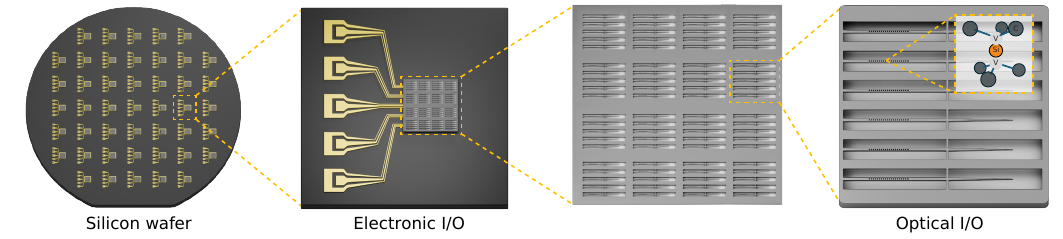}
\caption{Illustration of the scalable diamond quantum memory platform. A wafer-scale matrix of chips functionalized with electronic input-output (I/O) control lines such as coplanar waveguides for spin control of silicon-vacancy (SiV) quantum memories. Diamond membranes are deposited onto each individual chip and a matrix of nanophotonic devices is created using standard top-down nanofabrication techniques. High-quality photonic crystal cavities and tapered waveguide couplers provide a highly efficient photonic I/O interface to the optical transitions of embedded SiV quantum memories, enabling deterministic spin-photon interactions.}\label{fig1}
\end{figure*}
\section*{Introduction}\label{sec1}
Leveraging wafer-scale fabrication of semiconductor materials has been critical to the rapid improvements of classical computing and telecommunications technologies over the past 40 years. Many quantum technologies ranging from computing to networking to sensing would similarly benefit in both performance and scale from high-quality, high-yield wafer-scale fabrication\,\cite{awschalom_development_2021}. One example use-case for wafer-scale fabrication of quantum devices is quantum memories. Memories can be used to connect many individual quantum processing units via interconnect channels utilizing quantum memories as network buffer to compensate for channel losses\,\cite{bhaskar_experimental_2020}. The scalability of this approach hinges on the ability to fabricate large numbers of high-quality quantum memories, making solid-state memory approaches of particular interest. However, wafer-scale fabrication approaches are only directly applicable to a relatively small set of materials, preventing many materials hosting some of the most promising quantum memories from being produced at scale in an economically viable manner\,\cite{pompili_realization_2021, ruskuc_multiplexed_2025}.

Synthetic diamond is one example material which has long been recognized for its unique optical, quantum, and mechanical properties, but cannot be produced at wafer scale with sufficiently low strain and defect density for quantum technology applications. This fabrication bottleneck has slowed the deployment of diamond-based technologies for quantum sensing, computation, and communication, which these unique material properties enable\,\cite{awschalom_quantum_2018}.

In particular, diamond is among the leading platforms for the implementation of quantum sensors\,\cite{rondin_magnetometry_2014, degen_quantum_2017} and the production of efficiently interfaced, heralded quantum memories which are key to practical implementations of quantum networks as they enable repeat-until-success protocols\,\cite{azuma_quantum_2023,wehner_quantum_2018}. Optically active spin defects based on group-IV vacancy centers in diamond stand out for combining highly coherent optical interfaces and long spin coherence times\,\cite{rosenthal_single-shot_2024,sukachev_silicon-vacancy_2017} with the crucial advantage of inversion symmetry that minimizes spectral diffusion in nanophotonic devices\,\cite{evans_narrow-linewidth_2016}. The performance of a quantum network node depends not only on per-channel storage fidelity, efficiency, and bandwidth, but also on the total number of parallel memory channels\,\cite{ruskuc_multiplexed_2025}. Large-scale production of multi-channel diamond memory devices would mark a key step towards modular architectures for utility-scale quantum computers, and novel applications such as blind quantum computing\,\cite{wei_universal_2025}, long-baseline telescopes\,\cite{gottesman_longer-baseline_2012, khabiboulline_optical_2019} and distributed sensing\,\cite{proctor_multiparameter_2018, guo_distributed_2020}.

Fabrication of high-quality diamond photonic devices has long been accomplished using bespoke nanofabrication techniques on small, standalone bulk diamond dies which limits the yield and scale of the produced technologies. An angled-etching technique\,\cite{burek_free-standing_2012, stas_robust_2022, knaut_entanglement_2024} has been used to carve out devices from electronic grade bulk diamond, enabling proof-of-concept demonstrations of a variety of quantum optic and nanomechanical implementations. Unfortunately, these bulk-carving techniques are incompatible with foundry manufacturing of photonic integrated circuits. Quasi-isotropic diamond undercutting\,\cite{kuruma_controlling_2025, joe_observation_2025, rugar_quantum_2021, wan_large-scale_2020, khanaliloo_single-crystal_2015} has also been used to produce individual high-quality devices, but scaling this approach remains challenging due to roughness-limited quality factors and strong crystallographic and aspect ratio dependent etch behavior.

Here, we demonstrate a major advancement towards large-scale production of quantum memories with a deterministic, highly efficient optical interface, and heterogeneous integration to substrates functionalized with electronic control lines (Fig.\,\ref{fig1}).
Our approach involves high-yield deposition and parallelized patterning of a matrix of diamond thin films on a wafer featuring several pre-patterned individual dies. This technology represents a critical step towards scalable fabrication of hundreds of diamond devices in a single fabrication run on wafer-scale substrates without requiring access to wafer-scale quantum-grade diamond.

Our platform utilizes homogeneous, single-crystalline diamond films created by combining an ion implantation and liftoff technique with high-purity diamond overgrowth\,\cite{guo_tunable_2021, aharonovich_homoepitaxial_2012}. Such thin-film diamonds have been shown to be suitable for hosting quantum memories with lifetime-limited optical transitions and long spin coherence times\,\cite{guo_direct-bonded_2024}. Additionally, key quantum memory components including photonic interfaces, such as photonic crystal cavities\,\cite{ding_high-q_2024} and slow-light waveguides\,\cite{ding_purcell-enhanced_2025}, as well as spin control\,\cite{guo_microwave-based_2023} have been demonstrated in individual diamond films.

To demonstrate the large-scale fabrication capabilities enabled by our diamond films, we develop a robust fabrication pipeline for high-quality devices with close-to-unity yield. We report significant progress along several axes including high-quality, high-uniformity diamond overgrowth, deterministic implantation of color centers through nanoapertures, parallelized electrochemical etching and single-step millimeter-scale membrane deposition achieved via metal thermocompression bonding with near-unity yield. We demonstrate the deposition of tens of membranes to the same handling chip which can be extended to full wafer reconstitution where many individual membranes are tiled to cover the wafer surface for further processing.
\begin{figure*}[hbt]
\centering
\includegraphics[width=\textwidth]{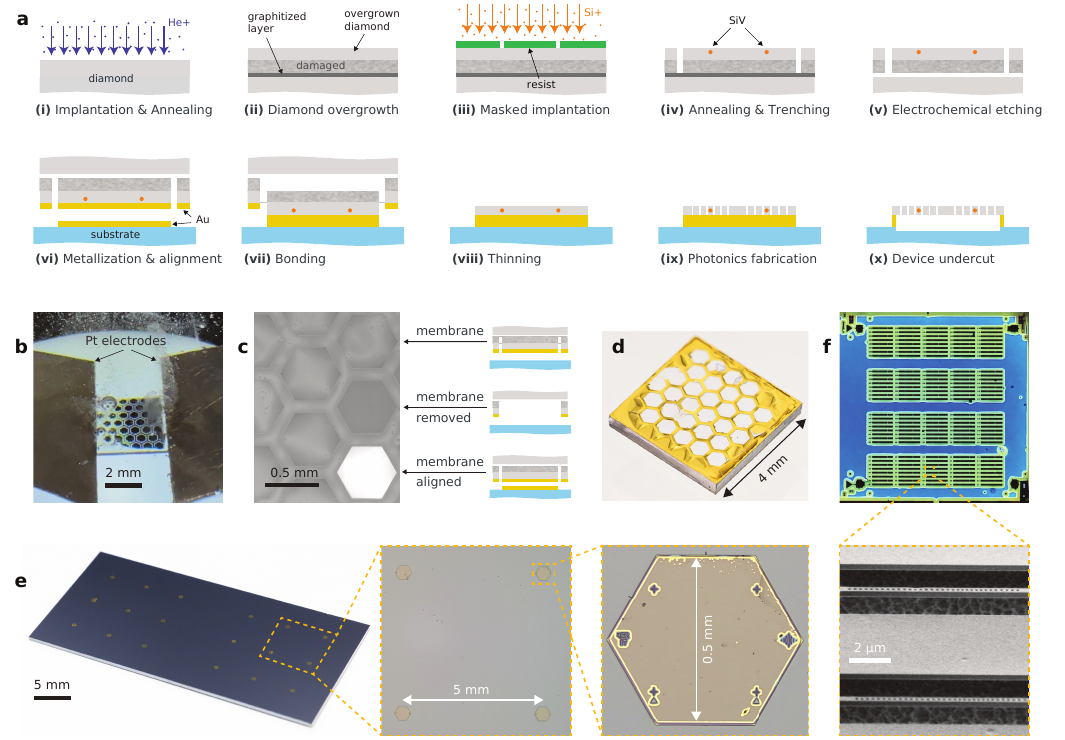}
\caption{\textbf{(a)} Fabrication process flow for diamond photonic devices (i) A graphitized layer at a well-defined depth is generated by a combination of high-energy He implantation and high-temperature annealing (ii) followed by high-purity diamond overgrowth. (iii) SiV centers are created by implanting Si at predetermined locations using nanoapertures and (iv) subsequent high-temperature annealing. A tethered membrane array is etched into the diamond (v) enabling access to the graphitized layer for removal via electro-chemical etching. (vi) The diamond surface is metalized (vii) and individual membranes are flip-chip bonded to a silicon substrate with commensurate metalized bonding pads. (viii) The membranes are thinned down removing the damaged diamond layer. (ix) Photonic crystal cavities are fabricated aligned to the implanted SiV centers and (x) devices are undercut via a combination of metal etchant and hydrofluoric acid. \textbf{(b)} Electro-chemical etching of the sacrificial graphitized layer. \textbf{(c)} Optical micrograph of the aligned flip-chip bonding process. Images of a metalized diamond membrane array and a commensurate bonding pad on the substrate are overlaid, facilitating precise alignment. Individual membranes are deposited sequentially by moving the substrate and aligning to individual membranes in the array. \textbf{(d)} Parent diamond die after transfer of 25 membranes. \textbf{(e)} Polarized photograph of a reconstituted $3\times6$ matrix of hexagonal membranes (0.5\,mm) arranged in a grid with 5\,mm spacing created by 18 consecutive transfers from the diamond shown in (d). Zoomed images were taken with a microscope. \textbf{(f)} Optical micrograph of an array of fabricated photonic crystal cavities and scanning electron micrograph of two representative photonic crystal cavity devices.}\label{fig2}
\end{figure*}

High-cooperativity interfaces for deterministic silicon-vacancy (SiV) memory-photon interactions can be implemented routinely on this platform. We demonstrate that we can fabricate photonic crystal cavities reliably and in parallel for several membranes bonded to the same handling chip with small spread in resonance wavelength while maintaining high quality factors.

Our process is streamlined by separating fabrication of optical components in diamond and functionalizing layers in the substrate. Device applications and performance can be tuned by introducing layers featuring different electronic or photonic components. Importantly, the high temperature processing necessary for creating defect centers is separated from front-end fabrication of electronic or photonic components with lower temperature tolerances (e.g. electronic control lines). Moreover, alignment between the substrate and devices on the film can be accomplished lithographically, eliminating the need for nanometer scale alignment during a pick-and-place integration step\,\cite{wan_large-scale_2020}.

When combined with wafer-scale fabrication, our approach paves the way towards large scale integration of memory-based quantum interconnect nodes with many efficient photonic and electronic channels. Our platform represents a crucial step towards utility-scale quantum computers implemented with efficient interconnection of individual quantum processing units based on ions\,\cite{chen_benchmarking_2024} or atoms\,\cite{bluvstein_logical_2024}.

\section*{Results}\label{sec2}
\subsection*{High-yield fabrication of high-quality diamond membranes}\label{subsec1}
We implement a high-yield, scalable fabrication pipeline for the generation of high-quality diamond membranes bonded to a semiconductor substrate. Our process combines ion implantation and membrane liftoff, high-purity diamond overgrowth, nanoaperture implantation of quantum memories and flip-chip thermocompression bonding to fabricate diamond membranes and photonic devices with high yield (Fig.\,\ref{fig2}(a(i-x)), methods).

We report several significant advancements compared to previously reported approaches\,\cite{guo_tunable_2021, aharonovich_homoepitaxial_2012}. First, by masking the edge of the diamond substrate during ion implantation, we retain a solid diamond frame which anchors the membrane during undercutting and bonding (Fig.\,\ref{fig2}(a,i)). Second, a systematic study of the growth conditions for homoepitaxial high-purity diamond results in consistent creation of diamond layers with low surface roughness ($<\,0.2\,\text{nm}$ root mean square (RMS)) and thickness variation ($<$20\,nm) across the diamond (Fig.\,\ref{fig2}(a,ii), methods). Finally, targeted ion implantation of SiV color centers through nanoapertures (Fig.\,\ref{fig2}(a,iii)) significantly increases the final device yield by reducing background fluorescence from unintentionally created, weakly coupled SiV centers.

To prepare an array of diamond membranes bonded to a wafer, we define a tethered array of micromembranes via aligned trench fabrication (Fig.\,\ref{fig2}(a,iv)) which enables collective undercutting of the whole membrane array simultaneously using electro-chemical etching (Fig.\,\ref{fig2}(a,v), Fig.\,\ref{fig2}(b)). We metalize the surface and use aligned flip-chip thermocompression bonding (Fig.\,\ref{fig2}(a,vi)) to deterministically transfer membranes from the parent diamond to handling chips with high throughput ($<1\,\text{min}$ per bond) and high yield (Fig.\,\ref{fig2}(a,vii)). Finally, photonic structures are fabricated by first thinning down the membrane to the desired thickness (Fig.\,\ref{fig2}(a,viii)), employing standard top-down semiconductor fabrication techniques (Fig.\,\ref{fig2}(a,ix)) and finally undercutting the structures using wet chemical etching (Fig.\,\ref{fig2}(a,x)).

\begin{figure*}[h]
\centering
\includegraphics[width=\textwidth]{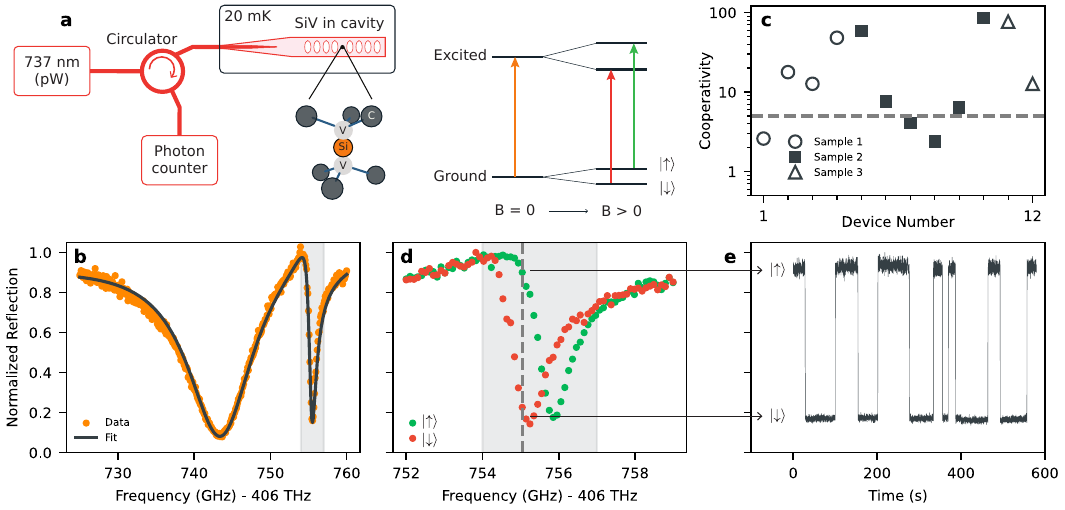}
\caption{\textbf{(a)} Schematic of the experimental setup and simplified level structure of an SiV center. \textbf{(b)} Laser scanning measurement of a representative cavity reflection spectrum. The cavity resonance is modulated by an SiV center coupled with a cooperativity of 92. \textbf{(c)} Statistics of the cooperativities of 12 SiVs in cavities measured over 3 chips. \textbf{(d)} Reflection spectrum showing spin-dependent device reflectivity. \textbf{(e)} Quantum jumps of the SiV spin under weak illumination.}\label{fig3}
\end{figure*}

Our high-throughput transfer process is a key step toward wafer-scale fabrication, enabling the creation of large-scale device arrays by depositing a matrix of many membranes on a single silicon handling chip without replacing the substrate or parent diamond. For this demonstration we define an array of hexagonal membranes with a lateral dimension of 0.5\,mm arranged in a honeycomb lattice weakly tethered to a frame. We employ gold thermocompression bonding to form a strong continuous bonding layer to a commensurate gold pad on a Si handling chip. When the parent diamond is retracted, the tethers break off the support frame leaving the membrane bonded. The next membrane is then bonded to the handling chip by moving to and aligning with the next bonding pad (Fig.\,\ref{fig2}(c)). The bonding process is repeated up to 25 times with a single parent diamond die (Fig.\,\ref{fig2}(d)). The malleable metal interlayer mitigates the effects of particle contamination and surface irregularities, such as hillocks or pits, that can form during overgrowth.
Figure\,\ref{fig2}(e) displays a $3\times6$ grid of membranes spaced by 5\,mm which were deposited consecutively showcasing the near-unity yield of our process.

\subsection*{Implementation of high-cooperativity interface to quantum memories}
Following membrane bonding to substrate, we create a deterministic, low-background photonic interface for diamond quantum memories by precisely aligning the fabrication of photonic crystal cavities with implanted color centers. After bonding, membranes are thinned to a thickness of 160\,nm resulting in a low surface roughness of less than $0.3\,\text{nm}$ RMS. We then fabricate a $4 \times 32$ array of photonic crystal cavities using standard top-down fabrication methods (Fig.\,\ref{fig2}(f), methods).

Our devices implement high-cooperativity interfaces for controlling SiV color centers in diamond, showcasing the excellent quality of our material. For characterization we measure the cavity reflection spectra using a movable tapered fiber probe in a dilution refrigerator with a base temperature of 20\,mK (Fig.\,\ref{fig3}(a)). We deposit nitrogen ice on the devices to red shift the cavity resonances and subsequently use a laser for controlled nitrogen sublimation to tune the cavities into resonance with embedded SiV centers\,\cite{nguyen_quantum_2019}. Scanning a narrow-linewidth laser we resolve sharp spectral features that reveal SiV centers which are well-coupled to the cavity mode (Fig.\,\ref{fig3}(b)). By fitting the modulation of the cavity reflection spectrum\,\cite{bhaskar_experimental_2020}, we extract a cavity cooperativity $C = 4g^2/ (\kappa_\text{tot} \cdot \gamma_\text{SiV}) = 92$. The SiV center at $f_\text{SiV} = 406.74\,\text{THz}$ is assumed to have a lifetime-limited linewidth $\gamma_\text{SiV}= 94\,\text{MHz}$, while the cavity has total linewidth of $\kappa_\text{tot} = 5.73\,\text{GHz}$. The atom-photon coupling strength is fitted to be $g = 3.52\,\text{GHz}$.

Figure\,\ref{fig3}(c) shows the distribution of cooperativities of devices which we pre-select via confocal PL measurements in a 4\,K cryostat across 3 different chips. Of the 12 devices measured, 9 featured SiV centers coupled with cooperativities $C > 5$, demonstrating that we can deterministically create efficient spin-photon interfaces for which the off-resonant cavity reflection is high and the reflection contrast between spin states is large and robust against spectral diffusion.

Next, we apply an external magnetic field along the SiV axis to lift the degeneracy of the SiV spin states (see level scheme in Fig.\,\ref{fig3}(a)). The exact alignment of the magnetic field will determine the excited state decay branching ratio of SiV into the spin states. Due to different g-factors in the ground and excited states, the optical transitions for different spin states are at different frequencies, resulting in spin-state-dependent cavity reflection (Fig.\,\ref{fig3}(d)). Locking the laser to the point of maximum reflection contrast between the two spin states, we observe quantum jumps from which we infer a spin-relaxation time of $T_1=32\,\text{s}$ under weak illumination setting a lower bound for the actual $T_1$ time (Fig.\,\ref{fig3}(e)).

\subsection*{Parallel fabrication of photonic devices}
\begin{figure}[ht]
\centering
\includegraphics[width=0.5\textwidth]{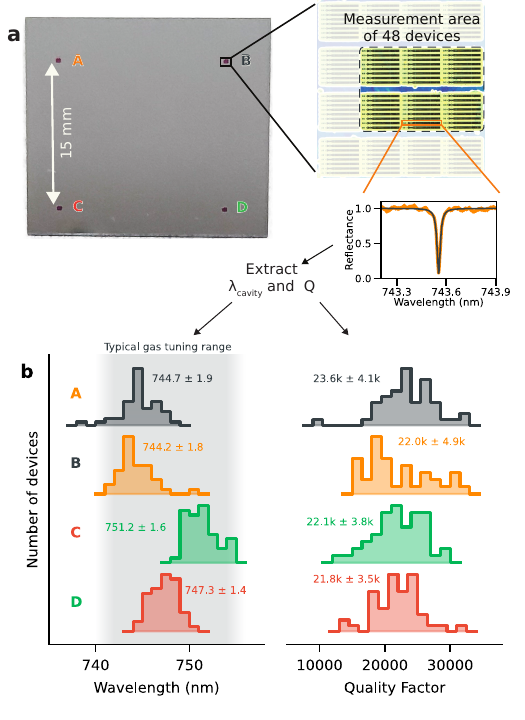}
\caption{\textbf{(a)} Bonding of 4 diamond membranes on the same handling chip spaced by 15\,mm for parallel fabrication of arrays of thin-film devices; Zoom on membranes after fabrication of a $4\times32$ array of photonic crystal cavities; Representative cavity reflection scan yielding a quality factor $Q=3.20\times10^4$ at wavelength $\lambda_\text{cav}= 743.5\,\text{nm}$. \textbf{(b)} Histograms of cavity resonance wavelengths and quality factors on the same block of 48 devices across the 4 different membranes.}\label{fig5}
\end{figure}

To confirm the compatibility of our platform with wafer-scale production, we fabricate parallel arrays ($4 \times 32$) of photonic crystal cavities of an identical design on four diamond membranes which were bonded 15 mm apart on a single handling chip (Fig.\,\ref{fig5}(a)). We measure quality factors and resonance wavelengths to capture differences in device performance introduced by variations in the membranes and the fabrication process. We focus on measuring a subset of 48 devices on every membrane (Fig.\,\ref{fig5}(a)) to limit the number of required measurements and exclude devices with known proximity effect shifts.

Figure\,\ref{fig5}(a) displays a reflection measurement for a representative device with quality factor $Q= 3.20\times10^4$ at wavelength $\lambda_\text{cav}= 743.5\,\text{nm}$ which is close to being critically coupled with a contrast of $92.5\%$. We determine quality factors and resonance wavelengths of our devices at room temperature by coupling a supercontinuum laser into the diamond waveguide through a tapered optical fiber interface and record a broadband reflection spectrum using a spectrometer.

We confirm the high quality of our devices and reproducibility of the fabrication by demonstrating average quality factors exceeding $2\times10^4$ with a small spread of resonance wavelengths $< 2\,\text{nm}$ across all fabricated membranes (Fig.\,\ref{fig5}(b)). We find wavelength distributions of $(744.7 \pm 1.9)\,\text{nm}$, $(744.2 \pm 1.8)\,\text{nm}$, $(751.2 \pm 1.6)\,\text{nm}$ and $(747.3 \pm 1.4)\,\text{nm}$ with corresponding $Q_\text{tot}$ factor distributions of $(2.36 \pm 0.41) \times10^4$, $(2.20 \pm 0.49) \times10^4$, $(2.21 \pm 0.38)\times10^4$, $(2.18 \pm 0.35)\times10^4$, respectively. The achieved spread in cavity wavelengths is well within the $\sim10\,\text{nm}$ cryogenic tuning range afforded by the employed nitrogen gas condensation technique, enabling a near-unity yield of usable, resonant cavities across an entire wafer. This is a significant advancement compared to angled etched devices, where the wavelength spread within one chip is so large that typically only $\sim~10\%$ of fabricated cavities are within the tuning range of optical transition of the SiV center.

We note that global shifts in wavelengths can easily be achieved by adjusting the input design. Run-to-run variations can be compensated by adding witness chips which are readily available due to the high-throughput of our membrane fabrication technique. Variations between membranes can be further improved by adjusting the input design based on metrology steps during fabrication to compensate for thickness variations of the diamond membrane, lithography resist and etching hardmask \,\cite{xin_wavelength-accurate_2025}. Overall, device optimization for diamond photonics based on previous techniques is typically slow or costly since optimization runs either require proxy fabrication tests or a large number of samples. After removing the material bottleneck and paving the way towards wafer-scale fabrication, we anticipate that in dedicated foundry environments the material platform will be improved further and will have close-to-unity yields.

\subsection*{Integration of membranes with functionalized substrates}

\begin{figure}[ht]
\centering
\includegraphics[width=0.5\textwidth]{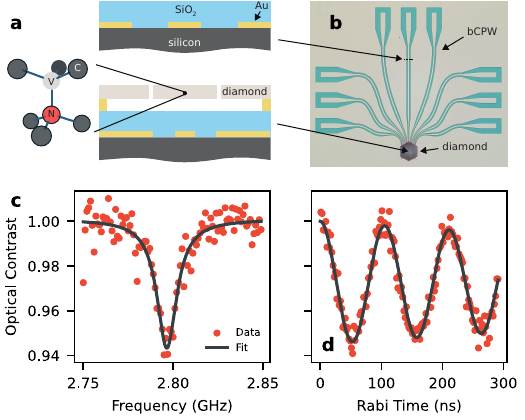}
\caption{\textbf{(a)} Schematic cross section of a buried coplanar waveguide (bCPW, top) and of a diamond membrane bonded above the bCPW (bottom). The Au bonding layer is partially removed via metal etching through access holes fabricated into the membrane enabling efficient spin driving of implanted NV centers. \textbf{(b)} Overview of a representative bCPW sample with 9 independent control lines. \textbf{(c)} Optically detected magnetic resonance and \textbf{(d)} Rabi oscillations of NV centers driven via the bCPW for a sample similar to the one shown in (b) where the membrane bonding pad is partially undercut to remove all bonding layer metal between a bCPW line and the diamond membrane.}\label{fig4}
\end{figure}

We showcase the potential for implementing efficient, fully packaged diamond quantum memories using our fabrication technology by demonstrating integration with functionalized substrates and highly efficient photonic packaging based on our previously demonstrated tapered-fiber technology\,\cite{burek_fiber-coupled_2017,zeng_cryogenic_2023}. We deposit diamond films implanted with nitrogen-vacancy centers onto silicon substrates functionalized with a buried coplanar waveguide (bCPW) for microwave delivery featuring a 1\,\textmu m thick silicon dioxide cladding layer (Fig.\,\ref{fig4}(a,b)). These substrates are fabricated at wafer scale and planarized using chemical mechanical polishing, enabling high-throughput and high-yield production. Diamond films can be bonded onto multiple different coplanar waveguide devices in parallel and then patterned collectively, further improving the rate at which integrated diamond photonic devices, such as sensors, can be produced. 

As a proof-of-concept, we drive spin transitions of implanted NV centers (methods) using microwave pulses generated via the bCPW. By adjusting an external magnetic field, we isolate and measure optically detected magnetic resonance of a small ensemble of NV centers with the same orientation (Fig.\,\ref{fig4}(c)). Coherent spin control is implemented by applying resonant microwave pulses through an integrated coplanar waveguide beneath the diamond membrane (Fig.\,\ref{fig4}(d)). We measure Rabi oscillations of the NV spin under continuous driving. A microwave power of 15\,dBm results in $t_\pi = 55\,\text{ns}$. With such microwave field strengths per input power the bCPW should immediately be applicable for the implementation of integrated SiV quantum memory devices. Importantly, the bCPW enables addressing of several devices in parallel by significantly reducing the heat load caused by microwave driving in contrast to patterning the CPW on the diamond surface\,\cite{psiquantum_team_manufacturable_2025}.

Harnessing modular fabrication and the potential for integration of diamond thin films with functionalized substrates constitutes a major step towards practical applications of quantum memories.
In previous approaches the fabrication of on-chip CPWs for microwave delivery and spin control required subsequent fabrication steps after the photonics fabrication was complete which increases complexity and reduces device yield. Moreover, aligned lithography and evaporation steps have an inherent risk of damaging devices due to additional handling steps. Furthermore, addressing several devices in parallel is challenging as heat is more difficult to dissipate for metal traces fabricated directly atop a diamond thin film.

\begin{figure}[ht]
\centering
\includegraphics[width=0.5\textwidth]{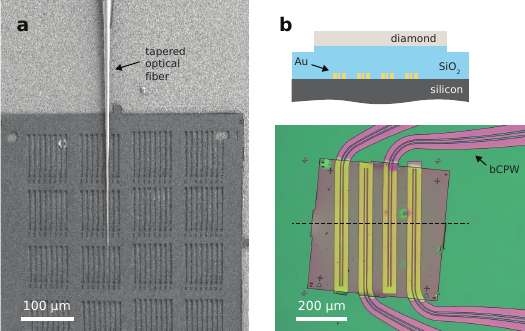}
\caption{\textbf{(a)} Scanning electron microscope image of a packaged device where the tapered fiber is permanently attached to a photonic crystal device via a tapered waveguide coupler.
\textbf{(b)}
Implementation of oxide bonding on bCPW chips to demonstrate the feasibility of a fully foundry-compatible approach without a metal interface layer.}\label{fig6}
\end{figure}

To further illustrate the scalability of our approach we demonstrate efficient optical packaging of thin-film devices which enables device operation without any active alignment. Figure\,\ref{fig6}(a) displays a scanning electron microscope image of a packaged device where the tapered fiber is permanently attached to a photonic crystal device via a tapered waveguide coupler. This fiber attachment method maintains a fiber-to-chip coupling efficiency comparable to that achieved through optimized active alignment techniques. The coupling efficiency of our system is measured to be $\eta_c \sim 85\%$ ($<1\,\text{dB}$ loss) by comparing the cavity reflection with that of a retroreflector. With further optimization, the insertion loss can be reduced to $<0.5\,\text{dB}$ as demonstrated in angled-etched devices\,\cite{zeng_cryogenic_2023}.

\subsection*{Discussion and Outlook}
This work represents a substantial advance in the scalable fabrication of high-cooperativity quantum devices. The thin-film diamond platform combines high-yield membrane deposition, deterministic color center placement, high-uniformity photonic crystal cavity fabrication, and microwave control. These features have collectively enabled the routine realization of spin-photon interfaces with cooperativities $C > 5$ achieved across multiple chips, highlighting the material and process quality. Among these, our best devices achieved very high cooperativities with $C \sim 100$ and are comparable to state-of-the-art implementations across solid-state platforms\,\cite{bhaskar_experimental_2020,najer_gated_2019}.

We highlight that our ability to fabricate membrane arrays with near-unity yield and the scalable, robust fabrication of photonic crystal cavities with desired properties on several membranes in parallel pave the way to wafer-scale device fabrication. We report high quality factors ($Q \sim 2\times10^4$) across multiple bonded membranes and a narrow spread of cavity resonances within individual membranes ($<2\,\text{nm}$). We anticipate that wavelength shifts between membranes can be eliminated by both adding metrology steps during fabrication and by transferring our process from a shared user facility to a dedicated fabrication facility. The demonstrated yield and uniformity mark a critical leap from earlier bespoke techniques that suffered from limited scalability and reproducibility.

In addition to high-speed microwave control, functionalized substrates will provide multiple electrical signals for the control of SiV quantum memories, for example via strain tuning\,\cite{machielse_quantum_2019} and could enable more complex complementary metal–oxide–semiconductor circuits for signal processing compatible with cryogenic environments\,\cite{pauka_cryogenic_2021}. Our platform enables integration with bCPW structures, as demonstrated by coherent microwave control of NV centers in thin-film diamond. Such integration forms the basis for the large scale production of packaged quantum sensors with low cost, size, weight, and power characteristics\,\cite{kim_cmos-integrated_2019}. 

This integration with bCPWs, combined with our previously demonstrated efficient optical fiber packaging, enables fully packaged, electrically and optically addressable quantum memory architectures. The number of optical I/Os, which is limited by the diamond membrane size, can be improved significantly by harnessing heterogeneous optical integration with platforms like lithium niobate or silicon nitride.

The use of metal thermocompression bonding in our current platform offers robust membrane attachment and high-throughput processing. Heterogeneous integration with both photonic and electronic components will be enabled by oxide bonding as demonstrated on bCPW substrates (Fig.\,\ref{fig6}(b)). This metal-free integration approach shows promise for simplifying downstream processing, improving cleanliness at the bond interface, and expanding material compatibility. We anticipate that in foundry-level fabrication the bonding yield will approach similar levels as with metal thermocompression bonding.

In conclusion, our work constitutes a major step toward wafer-scale, industrial-grade fabrication of integrated quantum memories and sensors. The demonstrated yield, performance, and versatility of our platform represent key enablers for building scalable quantum networks and sensing systems. We anticipate that continued development along the oxide bonding pathway and integration with photonic circuits will position thin-film diamond as a central material in future heterogeneous quantum technologies.

% comment to exclude from word count
%TC:ignore
\section*{Methods}
\subsection*{Diamond membrane fabrication}
Diamond membranes are fabricated by combining an ion implantation and liftoff technique with high-purity diamond overgrowth. A single-crystal diamond is first polished, and a subsurface graphitized layer is created at a well-defined depth via high-energy helium implantation (0.5\,MeV, 7-degree tilt, $5 \times 10^{16}\,\text{cm}^{-2})$. We mask the edge of the diamond during implantation resulting in a solid diamond frame which anchors the membrane during undercutting and bonding. High-temperature annealing ($1300\,^{\circ}\text{C}$, $1\times 10^{-6}\,\text{Pa}$, 120\,min) is used to convert this damage layer into graphite. After annealing, fuming H$_2$SO$_4$ + KNO$_3$ is used to remove the graphitization on the surface. Then a high-purity layer of diamond is produced above it via homoepitaxial high-purity overgrowth. This layer is grown to have a typical thickness of $\sim 0.5\,\text{\textmu m}$ with low thickness variation across the diamond (typically $< 10\,\text{nm} / 1\,\text{mm}$, Ext. Fig.\,\ref{extfig1}(a)) and low surface roughness ($< 0.2\,\text{nm}$ RMS, Ext. Fig.\,\ref{extfig1}(b)). Alignment marks are fabricated, and SiV color centers are introduced via aligned silicon ion implantation ($^{28}$Si, 125\,kV, $1\times 10^{12}\,\text{cm}^{-2}$) through nanoapertures in a PMMA/MMA resist stack and subsequent high-temperature annealing ($1250\,^{\circ}\text{C}$,$1\times 10^{-8}\,\text{Pa}$, 90\,min) enabling aligned integration with photonic devices. We then lithographically pattern trenches into diamond to define a tethered array of micromembranes. Finally, we undercut the membranes using electro-chemical etching\,\cite{tully_diamond_2021}. We define bonding pads on silicon substrates with respective sizes corresponding to the diamond micromembranes dimensions using optical lithography. We then deposit bonding material on both the surface of the diamond and the bonding pads utilizing either 50\,nm titanium and 100\,nm gold or silicon dioxide. Finally, we employ flip-chip bonding (Finetech Sigma) to deterministically bond the membranes with high yield at low temperature ($150\,^{\circ}\text{C}$), pressure (5\,kPa) and bonding time (15\,s). We deposit membranes with unity yield and high throughput ($< 1\,\text{min}$ per bond). To achieve the desired thickness suitable for photonic integration (typically between 100–300\,nm), the substrates were thinned via oxygen-based inductively coupled plasma reactive ion etching (ICP-RIE).

\subsection*{Fabrication of nanophotonic cavities}
We design single-sided photonic crystal cavities with a quadratically tapered lattice constant. The number of incoupling mirror units was optimized for preferential coupling to the waveguide mode, from which light is extracted using a tapered fiber probe. A silicon nitride (SiN) hard mask is deposited via plasma-enhanced chemical vapor deposition followed by spin-coating of ZEP520A resist. Photonic crystal cavities are defined by electron beam lithography and subsequently transferred to the SiN layer using inductively-coupled plasma reactive ion etching (ICP-RIE) with a C$_4$F$_8$/SF$_6$/H$_2$ gas mixture. After removal of the resist, the pattern is transferred into the thin-film diamond using O$_2$-based ICP-RIE. The SiN mask and the underlying metal bonding layer are then removed via immersion in dilute hydrofluoric acid and gold etchant and the devices are uniformly undercut via isotropic etching of the silicon substrate with XeF$_2$ 
\subsection*{Fabrication of buried coplanar waveguides}
bCPW samples are created on a 4-inch silicon wafer using metal liftoff and oxide deposition. We first pattern coplanar waveguides using photo lithography using a bilayer LOR/S1805 resist stack for liftoff and deposit 100\,nm gold with 10\,nm of chromium as adhesion layer on both sides. Next, we use inductively coupled chemical vapor deposition to deposit $\sim~2\,\text{\textmu m}$ of SiO$_2$ and apply chemical-mechanical polishing to smoothen the surface reducing the thickness to $\sim~1\,\text{\textmu m}$. NV centers were created in diamond membranes at a depth of $\sim~60\,\text{nm}$ under the surface by implanting nitrogen ions (N$^+$,50\,keV,$1\times10^{12}\,\text{cm}^{-2}$) and subsequent high-temperature annealing ($1250\,^{\circ}\text{C}$,$10\times 10^{-8}\,\text{Pa}$, 90\,min). The diamond membranes are thinned down to $\sim~300\,\text{nm}$ and the bonding pad is partially undercut by patterning holes into the diamond film to remove the bonding metal layer between membrane and bCPW lines. 
\subsection*{Experimental setup}
The experimental setup and methods used for mK nanophotonic cavity quantum electrodynamics experiments involving SiV centers is detailed in a separate publication\,\cite{nguyen_integrated_2019}. All measurements are conducted in a dilution refrigerator (BlueFors BF-LD250) with a base temperature of 20\,mK. This refrigerator includes a superconducting vector magnet (American Magnets Inc., 1-1-1\,T), a custom-built free-space wide-field microscope with a cryogenic aspherical lens, piezo positioners (Attocube ANPx101 and ANPx311 series), and both fiber and microwave feedthroughs. A gas condensation and laser backtuning technique is used to tune the nanocavity resonance. The SiV-cavity system is addressed optically through a fiber network via a tapered fiber coupler\,\cite{burek_fiber-coupled_2017,zeng_cryogenic_2023}, eliminating the need for free-space optics.

Experiments involving the microwave driving of NV centers via buried coplanar waveguides (bCPWs) were performed using a standard confocal microscopy setup. An ensemble of NV centers is optically excited using a green laser diode (Laes Electronics DLNSec 520) and the emission is read out using an avalanche photodiode (Excelitas SPCM 14/16) after being filtered through a 600 nm dichroic and a 670/70 nm bandpass filter. We collect APD counts using a timetagger (Swabian Time Tagger Ultra). In order to generate pulse sequences for microwave driving and gating of excitation/detection, we use a pulsestreamer (Swabian Pulsestreamer 8/2). We use a signal generator (R\&S SMB100B) as well as a microwave switch (CMCS0947A-C2) to generate the microwave signals required for ODMR and Rabi measurements.

\subsection*{Acknowledgments}
The authors thank Johannes Borregaard for helpful discussions, thank Maxwell Kleeman and Nicholas Mondrik for operational and infrastructure support and thank Mouktik Raha and Ankur Agrawal for contributions to the early exploratory phase of the project.

\setcounter{figure}{0}
\renewcommand{\figurename}{Extended Fig.}

\begin{figure*}[h]
\centering
\includegraphics[width=\textwidth]{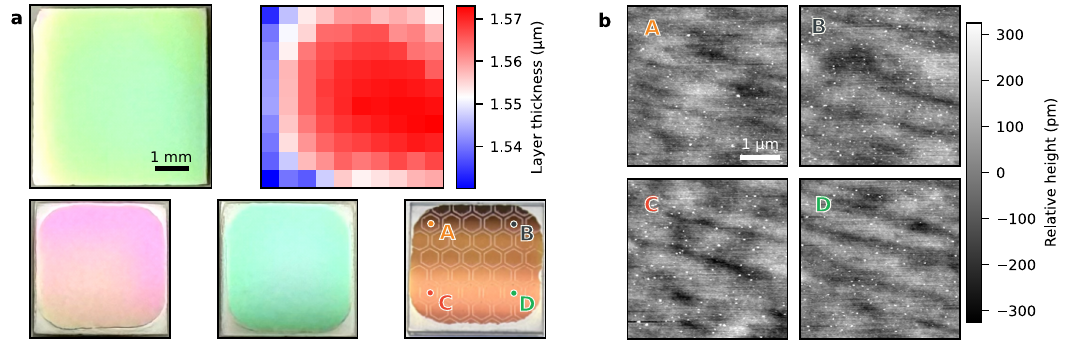}
\caption{\textbf{(a)} Top: Thickness variation of a diamond film on top of a sacrificial diamond layer for a $5.2\,\text{mm} \times 5.2\,\text{mm}$ diamond die recorded with (top left) a coherent white light source and (top right) using a Filmetrics F3 reflectometer revealing a film thickness of $(1.54 \pm 0.01)\,\text{\textmu m}$. Bottom: coherent white light source images of three additional samples with film thicknesses of $(1.38 \pm 0.01)\,\text{\textmu m}$, $(1.33 \pm 0.01)\,\text{\textmu m}$ and $(1.28 \pm 0.01)\,\text{\textmu m}$, respectively (from left to right). The scale bar of 1\,mm applies to every figure in the panel. \textbf{(b)} AFM measurements of the as-grown diamond surface for 4 distinct areas for the sample in panel (a,v) post trench fabrication to create the micromembrane array. The RMS roughness is $< 0.15\,\text{nm}$ across the sample. The scale bar of $1\,\text{\textmu m}$ applies to every figure in the panel.}\label{extfig1}
\end{figure*}

%TC:endignore
\bibliography{bib}% common bib file

\end{document}